\documentclass[usenatbib]{emulateapj}
\usepackage{graphicx}
\usepackage[flushleft]{threeparttable}
\usepackage[usenames,dvipsnames,svgnames,table]{xcolor}
\usepackage{hyperref}
\definecolor{darkblue}{rgb}{0.0,0.0,0.3}
\hypersetup{colorlinks,breaklinks,
            linkcolor=darkblue,urlcolor=darkblue,
            anchorcolor=darkblue,citecolor=darkblue}
\usepackage{amsmath,amssymb}
\usepackage{amsmath}

\begin{document}

\def\etal{et al.\ \rm}
\def\ba{\begin{eqnarray}}
\def\ea{\end{eqnarray}}
\def\etal{et al.\ \rm}
\def\Fdw{F_{\rm dw}}
\def\Tex{T_{\rm ex}}
\def\Fdis{F_{\rm dw,dis}}
\def\Fnu{F_\nu}
\def\FJ{F_J}

\newcommand\cmtrr[1]{{\color{red}[RR: #1]}}


\title{On the eccentricity excitation in post-main sequence binaries}

\author{Roman R. Rafikov\altaffilmark{1}}
\altaffiltext{1}{Institute for Advanced Study, 1 Einstein Drive, Princeton NJ 08540; 
rrr@ias.edu}


\begin{abstract}
Several classes of stellar binaries with post-main sequence (post-MS) components --- millisecond pulsars with the white dwarf companions (MSP+WD) and periods $P_b\sim 30$ d, binaries hosting post-asymptotic giant branch (post-AGB) stars or barium stars with $P_b\sim$ several yr --- feature high eccentricities (up to $0.4$) despite the expectation of their efficient tidal circularization during the AGB  phase. It was suggested that the eccentricities of these binaries can be naturally excited by their tidal coupling to the circumbinary disk, formed by the material ejected from the binary. Here we critically reassess this idea using simple arguments rooted in the global angular momentum conservation of the disk+binary system. Compared to previous studies we (1) fully account for the viscous spreading of the circumbinary disk, (2) consider the possibility of reaccretion from the disk onto the binary (in agreement with simulations and empirical evidence), and (3) allow for the reduced viscosity after the disk expands, cools, and forms dust. These ingredients conspire to significantly lower the efficiency of eccentricity excitation by the disk tides. We find that explaining eccentricities of the post-AGB binaries is difficult and requires massive ($\gtrsim 10^{-2}M_\odot$), long-lived ($\gtrsim 10^5$ yr) circumbinary disks, that do not reaccrete. Reaccretion is detrimental for the eccentricity growth also in the MSP+WD systems. Reduced efficiency of the disk-driven excitation motivates study of alternative mechanisms for producing the peculiar eccentricities of the post-MS binaries.   
\end{abstract}




\section{Introduction}  
\label{sect:intro}


Post-main sequence (post-MS) stellar evolution in binaries is known to often result in significant modification of their orbital parameters. Systems, which are compact enough, can easily reach semi-detached stage with the Roche lobe overflow (RLOF), during which mass can be exchanged between the components. During this phase semi-major axis of the binary evolves \citep{Thomas}, possibly in a dramatic fashion if e.g. binary enters the common envelope phase \citep{Iben}. Mass exchange can naturally produce unusual abundance patterns in the binary components \citep{McClure83,McClure84}. Moreover, tides in the evolved star are thought to damp eccentricity of the system to very small value \citep{Zahn,Shu}.  

This expectation was challenged by the discovery of a number of post-asymptotic giant branch (post-AGB) stars, many of them extremely deficient in iron, to reside in binaries with eccentricities as high as 0.3 and orbital periods $P_b\sim 1$ yr \citep{vanWinckel1995}. At such separations the binary should have been circularized by the tides during the AGB phase of the evolved companion. Moreover, a population of stars with strong barium lines in their spectra (Ba stars) was also found to reside in eccentric binaries with periods $\lesssim 10^4$ d. Their binarity naturally explains \citep{Jorissen} the overabundance in the s-process elements as being a result of the past accretion of mass (most likely via dense wind) from the progenitors of their more evolved companions, thought to be the white dwarfs (WDs) based of the mass function measurements \citep{McClure1990}). But the high eccentricities of these systems (up to 0.3-0.4) at orbital periods $\lesssim 10$ yr are difficult to understand, given that tides are expected to circularize them for $P_b\lesssim 3000$ d during the AGB phase \citep{Pols}.  

Recently, another class of evolved, short-period, and anomalously eccentric binaries was found. Several binary millisecond pulsars (MSPs) with the WD companions and orbital periods clustered around $P_b\approx 30$ d were found to have eccentricities in the range $0.03-0.13$ \citep{Antoniadis2016}. While they are not as non-circular as the post-AGB and Ba star-hosting binaries, their eccentricities are still three orders of magnitude higher than the expectation for the MSP-WD binaries with $P_b\approx 30$ d following the RLOF phase \citep{Phinney,PhinneyKulkarni}.  

Existence of these different classes of the evolved eccentric binaries calls for a mechanism that could increase binary eccentricity starting from a circular orbit {\it during or after} the AGB phase. The ideas for such a mechanism explored in the past include the enhanced mass loss and transfer during the periastron passages \citep{Soker,Bona} and kicks received by the WD at birth \citep{Izzard}.

More recently it was suggested \citep{Dermine,Antoniadis} that the eccentricities of evolved binaries can be pumped up as a result of their gravitational interaction with the massive circumbinary disk. Indeed,  \citet{Lubow1996} and \citet{Lubow2000} demonstrated that the tidal coupling between the circumbinary disk and the binary should lead to eccentricity growth, in agreement with numerical simulations of the circumbinary disks \citep{Cuadra2009,Farris2014}. Observational motivation for this possibility lies in numerous detections of dusty disks around post-AGB stars \citep{vanWinckel2006,Gielen,vanAarle}. Such a disk can naturally form as a result of the mass outflow through the L2 Lagrange point in the Roche lobe-filling binary or from the dense AGB wind \citep{Blundell,Pejcha}. In the case of MSPs with the WD companions it was proposed by \citet{Antoniadis} that the disk would form from the material escaping the WD progenitor as it undergoes hydrogen shell flashes prior to entering the WD cooling phase. \citet{Dermine} and \citet{Antoniadis} calculated eccentricity growth driven by tidal coupling of the binary to a non-evolving circumbinary disk with mass $M_d\sim 10^{-4}-10^{-2}M_\odot$ and found reasonable agreement with observations. 

However, such low mass disks have very little capacity to absorb the angular momentum of the binary, which is needed for raising its eccentricity appreciably. Specific angular momentum of the disk formed by the L2 ejection is initially comparable to the specific angular momentum of the binary itself, meaning that a non-evolving low-$M_d$ disk cannot be very effective at increasing the binary eccentricity. To obtain binary $e\gtrsim 0.1$ the disk has to actually {\it expand} viscously, thus absorbing the binary angular momentum. However, this lowers its surface density, and thus {\it decreases} the amplitude of the torque acting on the binary with time. This is something that was not accounted for by  \citet{Dermine} and \citet{Antoniadis}.

In this work we revise the post-MS binary eccentricity excitation by properly considering the angular momentum exchange between the binary and a {\it viscously evolving} circumbinary disk. In addition to the disk evolution, we also account for the possibility of {\it reaccretion} of some disk material by the binary using the results of \citet{Rafikov2016}. This additionally lowers the strength of the disk-binary coupling. Previously, the reaccretion of the metal-poor gas from the circumbinary disk, in which the refractory component has condensed into dust grains, has been proposed \citep{Waters} to explain the extreme Fe depletion seen in many binary post-AGB stars \citep{vanWinckel1995,Maas}. Efficient accretion from the circumbinary disk is also seen in numerical simulations \citep{MacFadyen2008,Farris2014,ShiKrolik2015,Munoz}. Thus, reaccretion should be an important factor in the binary-disk coupling. As we show in this work, both disk evolution and reaccrition conspire to dramatically reduce binary eccentricity growth, putting in question the efficiency of the tidal mechanism of the eccentricity excitation. 

Our work is organized as follows. We describe our basic setup in \S \ref{sect:basic} and provide simple estimates of the eccentricity excitation in \S \ref{sect:simple}. Viscous evolution of the circumbinary disk is covered in \S \ref{sect:visc}, while the results for the binary eccentricity evolution are presented in \S \ref{sect:bin_ecc}. We discuss them in \S \ref{sect:disc}, focusing on the role of reaccretion (\S \ref{sect:reaccr}), validity of our assumptions (\S \ref{sect:validity}), and comparison with the existing work (\S \ref{sect:compar}). We summarize our results in \S \ref{sect:sum}.


\section{Basic setup}  
\label{sect:basic}


We consider a binary consisting of the primary and secondary with masses $M_p$ and $M_s$ (total mass is $M_b=M_p+M_s$) with the initial semi-major axis $a_b$. The binary is surrounded by a coplanar gaseous disk with initial mass $M_d(0)$, which forms next to it as a result of mass loss from the binary \citep{Pejcha}. For that reason the initial specific angular momentum of the disk material should be comparable to that of the binary itself. Thus, the starting total angular momentum of the disk $L_d(0)$ can be parametrized as $L_d(0)=\eta M_d(0)\left(G M_b a_b\right)^{1/2}$, where $\eta$ is a constant of order unity. 

Simulations of the circumbinary disks \citep{Farris2014,ShiKrolik2015,Munoz} invariably show the formation of a central cavity, in which the binary orbits. This cavity is cleared by the binary torque on the disk arising from the non-axisymmetric part of the binary potential. Its radius is typically found to be about $2a_b$ \citep{MacFadyen2008}. Despite this tidal barrier, matter from the inner edge of the disk is still seen to penetrate into the cavity and get accreted by the binary, which is very important for the issue of reaccretion discussed in \S \ref{sect:reaccr}. 

By assumption, the binary starts out circular as a consequence of the efficient tidal damping in the extended envelope of the evolved companion prior to that moment. After the disk forms, the binary eccentricity $e$ increases as a result of the gravitational coupling with the circumbinary disk. The same coupling also reduces semi-major axis of the binary, thus lowering its angular momentum  $L_b=\mu_b\left[G M_b^3 a_b\left(1-e^2\right)\right]^{1/2}$ in two ways. Here $\mu_b\equiv q/(1+q^2)$ is the dimensionless reduced mass of the system (normalized by $M_b$), and $q\equiv M_s/M_p<1$ is the mass ratio of the components. 

For a given change of the $L_b$ the highest change of the binary eccentricity is achieved if the semi-major axis of the binary stays constant in the course of the disk-driven orbital evolution. To set an upper limit on the disk-induced $e$ we will adopt this optimistic assumption and set $a_b=$ const throughout this work (see \S \ref{sect:validity}). This also simplifies calculation since all the disk torque then goes into the eccentricity excitation. 

The change of the binary angular momentum as its eccentricity increases from zero to $e$ is
\ba    
\Delta L_b(e)=\mu_b\left(G M_b^3 a_b\right)^{1/2}f(e),
\label{eq:dL}
\ea    
with $f(e)=1-\sqrt{1-e^2}$. In the small eccentricity limit $f(e)\approx e^2/2$.

Angular momentum shed by the binary gets absorbed by the circumbinary disk. The latter viscously expands out at the rate controlled by its internal stresses. It is this process that ultimately sets the rate, at which the binary loses its angular momentum, see \S \ref{sect:disc}. Here we will assume standard $\alpha$-ansatz for the kinematic viscosity \citep{shakura_1973}, $\nu=\alpha c_s^2/\Omega$, where $\Omega$ is the local angular frequency, $c_s=(k_B T/\mu)^{1/2}$, and $T$ is the midplane temperature of the disk. 

In this work we will neglect internal dissipation as the heating source of the disk and assume that $T$ is determined by the  central stellar irradiation. This is a reasonable expectation since the luminosities of the post-MS stars in this phase are high, $L_\star\gtrsim 10^3L_\odot$. Thermal balance under central illumination results in the following approximate relation for the temperature profile:
\ba     
T(r) &\approx &\left(\frac{\zeta L_\star}{4\pi\sigma}\right)^{1/4}r^{-1/2}
\label{eq:Tr}\\
&\approx &1.3\times 10^3~\mbox{K} ~\left(\zeta_{0.1}L_{\star,3}\right)^{1/4}r_1^{-1/2},
\ea   
where $L_{\star,3}\equiv L_\star/(10^3L_\odot)$. Constant factor $\zeta=0.1\zeta_{0.1}<1$ approximately accounts for the fact that the disk surface intercepts starlight at a grazing incidence angle. This generally results in $\zeta$ being a (weak) function of $r$ \citep{Chiang}, but in this work we will simply set $\zeta=0.1$.

It is easy to see that the viscous time $t_\nu=r^2/\nu$ in the disk with such temperature structure scales linearly with $r$. Moreover, beyond $\sim 1$ AU the gas would be molecular and cool enough to form dust, as is in fact observed in many post-AGB binaries \citep{vanWinckel2006,vanAarle}. This should dramatically reduce the effective viscosity in the disk. Making an analogy with the protoplanetary disks, which feature similar temperatures (albeit at somewhat smaller separations) and viscosity $\alpha\sim 10^{-3}-10^{-2}$ \citep{Hartmann}, in this work we will adopt $\alpha=10^{-2}$ as a typical viscosity value.


\section{Simple estimates}  
\label{sect:simple}


We now provide a simple order of magnitude estimate of the eccentricity that can be reached by the binary. We will focus on a rather optimistic situation, in which the binary {\it does not reaccrete} any gas from the disk, which requires that the central torque is strong enough to suppress any inflow into the central cavity. Then the disk mass is conserved and is equal to $M_d(0)$ at all times. Most of the mass is concentrated near the outer edge of the disk at the radius $r_{\rm out}$ that steadily grows with time as a result of viscous expansion. The angular momentum of this disk is $L_d(t)\sim M_d(0)\sqrt{G M_b r_{\rm out}(t)}$, which should equal $\Delta L_b$ when $r_{\rm out}\gg a_b$. Specializing to the low-$e$ limit and dropping constant factors one can then estimate $e\approx (M_d(0)/\mu_b M_b)^{1/2}(r_{\rm out}/a_b)^{1/4}$.

Outer scale $r_{\rm out}$ is the radius, at which the viscous time $t_\nu$ is equal to the time that elapsed since the start of the disk evolution. Equating $t_\nu=\alpha^{-1}\Omega(r_{\rm out})r_{\rm out}^2c_s^{-2}(r_{\rm out})$ to $t$ and using temperature profile (\ref{eq:Tr}) one finds that 
\ba    
r_{\rm out}(t)\approx 40~\mbox{AU}~\frac{t}{10^5\mbox{yr}}~\frac{\left(\zeta_{0.1}L_{\star,3}\right)^{1/4}}{M_{b,1}^{1/2}}\alpha_{-2},
\label{eq:r_out}
\ea      
where $M_{b,1}\equiv M_b/M_\odot$, $\alpha_{-2}\equiv\alpha/10^{-2}$, and we took $\mu=2m_p$.

Using this expression we can evaluate the binary eccentricity reached in time $t$ as
\ba      
e(t) & \approx & 0.06\left(\frac{t}{10^5\mbox{yr}}\right)^{1/4}\left(\frac{M_d}{10^{-4}M_\odot}\right)^{1/2}
\nonumber\\
& \times &\frac{\left(\zeta_{0.1}L_{\star,3}\right)^{1/16}}{a_{b,1}^{1/4}M_{b,1}^{5/8}}\alpha_{-2}^{1/4},
\label{eq:et}
\ea      
where $a_{b,1}\equiv a_b/$AU and we assumed $\mu_b=0.2$. Note the weak scaling of $e$ with many system parameters --- $L_\star$, $\alpha$, $a_b$, etc. The slow dependence of the eccentricity on time is a direct consequence of the viscous spreading of the circumbinary disk, as we will see next.


\section{Viscous evolution of the disk}  
\label{sect:visc}


\begin{figure}
\centering
\includegraphics[width=0.48\textwidth]{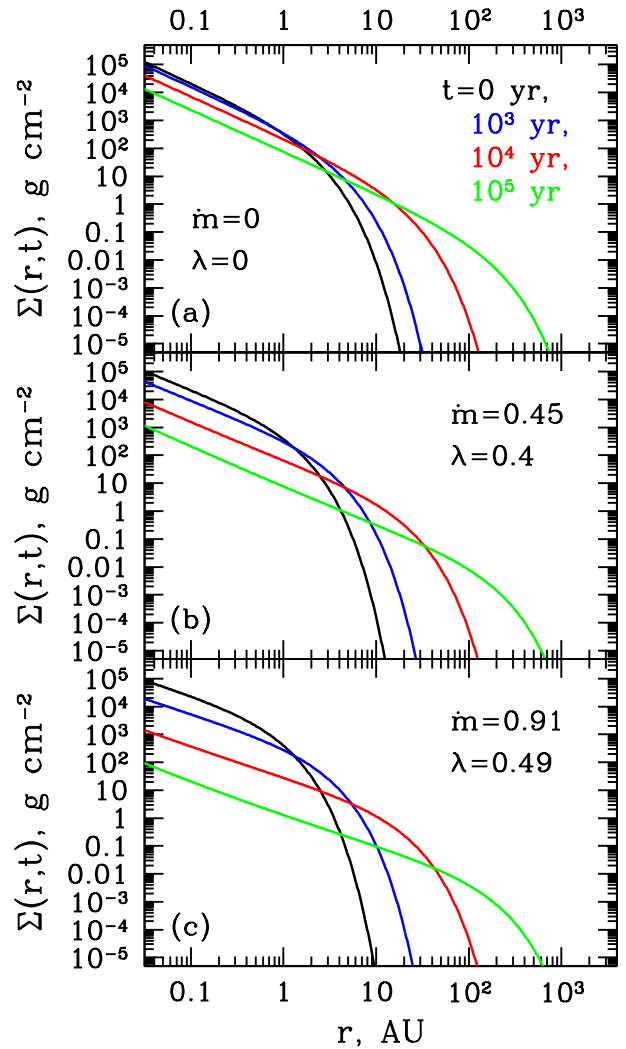}
\caption{Self-similar evolution of the disk surface density. Snapshots of $\Sigma(r,t)$ at different moments of time (indicated on top panel) are shown for different degrees of reaccretion: (a) $\dot m=0$, no accretion, (b) $\dot m=0.45$, significantly suppressed accretion, and (c) $\dot m=0.91$, reaccretion close to maximum value (realized in the case of zero central torque). Corresponding values of the similarity parameter $\lambda$ directly related to $\dot m$ \citep{Rafikov2016} are also indicated. Note the much faster decay of $\Sigma$ in the case of weakly suppressed reaccretion, resulting in the less efficient gravitational coupling with the binary.
\label{fig:selfsim}}
\end{figure}

 Now we examine viscous evolution of the circumbinary disk in more detail. Soon after its outer radius exceeds the semi-major axis of the binary $a_b$, the disk should start evolving in a self-similar fashion, independent of the initial conditions \citep{lynden-bell_1974}. At the same time, the structure of the expanding disk would still depend on the {\it boundary conditions} at the disk center imposed by the central object. This was previously demonstrated by \citet{lynden-bell_1974}, \citet{filipov_1984}, \citet{lyubarskij_1987}, \citet{Cannizzo}, who derived two kinds of the self-similar disk solutions for different inner boundary conditions: the one with no torque at the center and the one with no inflow (or outflow) in the disk center. 

Recently \citet{Rafikov2016} extended these self-similar solutions to disks with the {\it arbitrary} degree of reaccretion $\dot m$, which is defined as the ratio of the central $\dot M$ (reduced by the non-zero central torque) for a disk with given total mass $M_d$ and angular momentum $L_d$ to central $\dot M$ of the disk with the same $M_d$ and $L_d$ but zero central torque (i.e. no suppression of accretion). Previously known solutions with no central inflow correspond to $\dot m=0$, while the solutions with zero central torque and unsuppressed accretion naturally result in $\dot m=1$. New self-similar solutions found in \citet{Rafikov2016} are parametrized by a similarity parameter $\lambda$, which is uniquely related to $\dot m$, as shown in that work. 

We employ these solutions in our work and show in Figure \ref{fig:selfsim} the self-similar evolution of the surface density profile of a circumbinary disk for different values of the similarity parameter $\lambda$. This calculation assumes $T(r)$ given by equation (\ref{eq:Tr}), $\alpha=10^{-2}$, and initial disk mass $M_d(0)=10^{-3}M_\odot$. The outer disk radius (radius where the slope of $\Sigma$ starts going down) evolves in reasonable agreement with the simple estimate (\ref{eq:r_out}). One can see that as the disk viscously expands, gas density near the binary\footnote{Strictly speaking, these self-similar solutions apply only outside several$\times a_b$ but for simplicity we plot them as extending to $r=0$.} ($r\sim a_b=0.1$ AU) steadily goes down. This reduces the efficiency of the gravitational coupling between the binary and the disk over time, and weakens the angular momentum loss of the binary at late stages. The central density reduction is apparently slower and smaller for lower values of $\lambda$ (or $\dot m$). This is because lower $\lambda$ means higher degree of the accretion suppression at the disk center \citep{Rafikov2016}, so that more mass is retained in the disk as it evolves (disk mass is conserved for $\lambda=0$, when $\dot m=0$).

As shown in \citet{Rafikov2016}, temperature profile $T\propto r^{-1/2}$ corresponds to a particular viscosity behavior such that $\nu$ is independent of the local surface density $\Sigma$. We normally call this a {\it linear} viscosity behavior. More complicated cases of the {\it nonlinear} (i.e. $\Sigma$-dependent) behavior of $\nu$ can also occur when the internal dissipation plays important role in heating the disk \citep{Rafikov2016}, but they will not be considered here. 

\begin{figure}
\centering
\includegraphics[width=0.48\textwidth]{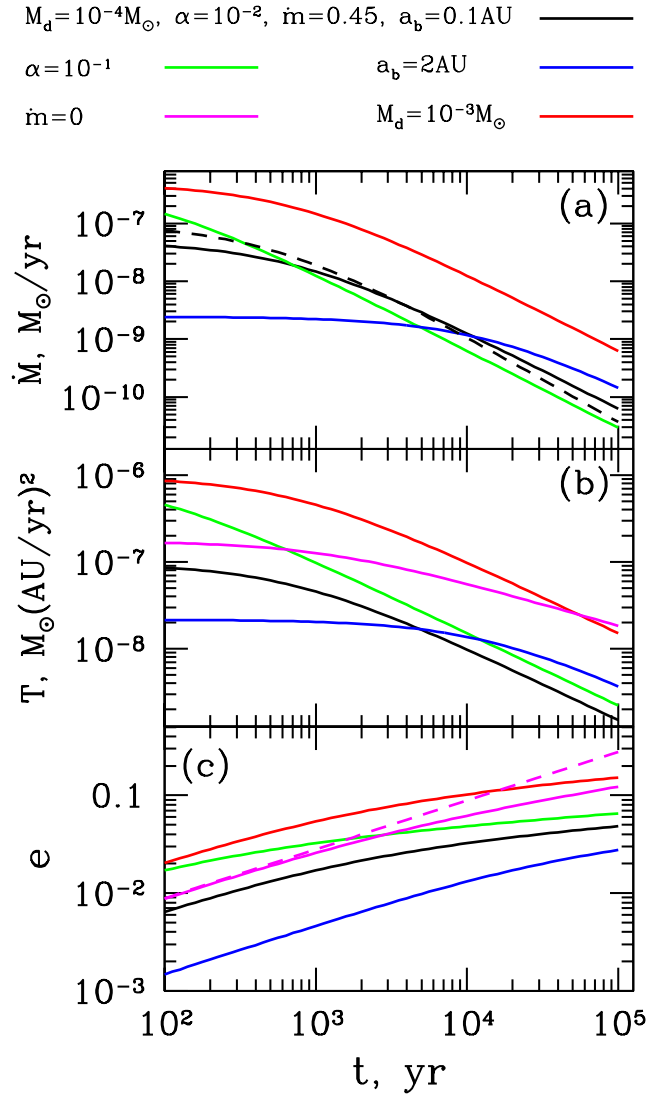}
\caption{Time evolution of the parameters of the gravitationally coupled binary and viscously evolving circumbinary disk: (a) central mass accretion rate, (b) central torque acting on the disk (equal to the rate at which binary loses angular momentum), (c) binary eccentricity $e$. Black curves correspond to the fiducial case of a disk with initial mass $M_d=10^{-4}M_\odot$, $\alpha=10^{-2}$, and partly suppressed accretion $\dot m=0.45$ (see middle panel of Figure \ref{fig:selfsim}) around a $a_b=0.1$AU, $M_b=1M_\odot$ binary with mass ratio $q=0.3$. Other curves show results for disks in which one of these parameters is varied as indicated in the legend on top. Black dashed curve in panel (a) is for $\dot m=0.91$ (see bottom panel of Figure \ref{fig:selfsim}). Dashed magenta curve in panel (c) is for $\dot m=0$ disk (no reaccretion), in which additionally the torque on the binary is always kept at its initial value (i.e. viscous disk spreading is {\it artificially disregarded}). 
\label{fig:time_ev}}
\end{figure}

As a result of viscous expansion and gravitational coupling to the binary the angular momentum of the disk $L_d$ grows with time. \citet{Rafikov2013} showed that 
\ba     
L_d(t) &=& L_d(0)\left(1+\frac{t}{k_\varphi t_0}\right)^{1-k_\varphi},~~~k_\varphi = \frac{1}{2(1-\lambda)},
\label{eq:L_devolution}
\ea     
where $t_0$ is the initial viscous time scale of the disk, discussed next. These expressions pertain to the case of a disk with $\alpha$ independent of $\Sigma$ and temperature profile (\ref{eq:Tr}). Also, as mentioned above, $\lambda$ is a monotonic function of the central accretion efficiency $\dot m$. For $\dot m=0$ (fully suppressed reaccretion) one finds $\lambda=0$ and $k_\varphi=1/2$. For $\dot m=1$ (free reaccretion, zero central torque) one gets $\lambda=1/2$ (for $T(r)\propto r^{-1/2}$, \citet{Rafikov2016}) and $k_\varphi=0$, i.e. disk does not gain any angular momentum from the binary.

Initial viscous time of the disk can be computed for the temperature profile (\ref{eq:Tr}) as  \citep{Rafikov2016} 
\ba     
t_0 &=& \frac{4}{3}\frac{\mu}{k_B}\frac{a_b}{\alpha}\left[\frac{4\pi\sigma (G M_b)^2}{\zeta L_\star}\right]^{1/4}\left(\frac{\eta}{I_L}\right)^2
\label{eq:t0}\\
&\approx & 1.5\times 10^4~\mbox{yr}~\frac{a_{b,1}}{\alpha_{-2}}\frac{M_{b,1}^{1/2}}{(\zeta_{0.1}L_{\star,3})^{1/4}}\frac{\eta_2^2}{I_L^2},
\ea      
where $\eta_2\equiv \eta/2$. A constant factor $I_L\sim 1$ characterizes spatial distribution of the angular momentum in the disk, and is a function of $\lambda$ as described in \citet{Rafikov2016}. Presence of this parameter, as well as other numerical factors in equation (\ref{eq:t0}), stem from the fact that the self-similar solutions in \citet{Rafikov2016} were formulated in terms of the viscous angular momentum flux and specific angular momentum\footnote{In the notation of \citet{Rafikov2016} a decretion disk with $\Sigma$-independent viscosity and temperature profile $T(r)\propto r^{-1/2}$ corresponds to the $d=0$, $p=0$ linear solution, for which a number of analytical results can be derived.}, rather than $\Sigma$ and $r$. Equations (\ref{eq:L_devolution})-(\ref{eq:t0}) are accurate as long as the disk structure is close to self-similar, which should certainly be the case for $t\gg t_0$.

Differentiating the expression (\ref{eq:L_devolution}) with respect to time one finds the central torque $T(t)$ acting on the disk
\ba    
T(t)=\frac{1-k_\varphi}{k_\varphi}\frac{L_d(0)}{t_0}\left(1+k_\varphi^{-1}\frac{t}{t_0}\right)^{-k_\varphi}
\label{eq:Ldot}
\ea     
Angular momentum conservation dictates that equal and opposite torque acts on the binary, increasing its eccentricity. It is important to realize that it is the global evolution of the disk (with the boundary conditions for accretion at its center imposed by the binary) that determines the torque on the binary. The binary torque (strength of the associated resonances) always self-adjusts to be equal to $T(t)$ \citep{Lubow1996}, by rearranging the disk surface density at resonant locations.

Time evolution of $T$ for different system properties is illustrated in Figure \ref{fig:time_ev}b. Our fiducial case features some reaccretion of the disk material, so that the binary accretes at the rate, which is about $45\%$ of the $\dot M$ that a disk with zero central torque would have. Other curves are produced by varying different system parameters: disk mass, viscosity, $\dot m$ and $a_b$.

Whenever $\dot m\neq 0$ central binary re-accretes some of the mass initially deposited in the circumbinary disk. Central mass accretion rate onto the binary is given by \citep{Rafikov2016}
\ba    
\dot M(t)=\lambda\frac{M_d(0)}{t_0}\left(1+k_\varphi^{-1}\frac{t}{t_0}\right)^{-(2-\lambda)k_\varphi},
\label{eq:Mdot}
\ea     
where $\lambda$ is directly related to $\dot m$ as described in \citet{Rafikov2016}. Evolution of $\dot M(t)$ is shown in Figure \ref{fig:time_ev}a for the system parameters corresponding to the central torque curves in Figure \ref{fig:time_ev}b (we do not show $\dot m=0$ curve as it has $\dot M=0$; to illustrate the role of reaccretion we show a black dashed curve for $\dot m=0.91$ instead). We provide more in-depth discussion of these profiles in \S \ref{sect:disc}.


\section{Binary eccentricity evolution}  
\label{sect:bin_ecc}


Growth of the binary eccentricity is determined by the conservation of the angular momentum of the coupled disk+binary system. Conservatively assuming that all angular momentum loss of the binary goes towards the growth of its eccentricity (i.e. $a_b\approx$ const), one arrives at the relation 
\ba   
\Delta L_b(e)=L_d(t)-L_d(0),
\label{eq:master}
\ea       
which allows one to determine $e(t)$ using equations (\ref{eq:dL}), (\ref{eq:L_devolution})-(\ref{eq:t0}).

Eccentricity evolution calculated in this fashion is shown in Figure \ref{fig:time_ev}c for different system parameters. Comparing different curves we find that binaries with more massive disks (red) reach higher $e$ than their counterparts with lower $M_d$ (black). Wider binaries (blue) exhibit considerably lower final $e$ than the short-period ones (black). Higher $\alpha$ (green vs. black) results in faster viscous evolution and higher final $e$. Also, the increase of $e$ at late times is very slow. All this is in perfect agreement with our estimate (\ref{eq:et}).

For comparison, we also show how the eccentricity would evolve if the disk were not viscously spreading. In this case we fix the central torque at its initial value $T(0)$ throughout the whole evolution. This calculation is done for $\dot m=0$ (fully suppressed accretion) and is shown by the dashed magenta curve in Figure \ref{fig:time_ev}c. It should be compared with the solid magenta curve, which was also computed for $\dot m=0$ but with properly evolved circumbinary disk. One can see that not accounting for the disk evolution overestimates the binary eccentricity excitation by almost a factor of 3 at $t=10^5$ yr. This clearly demonstrates that to understand binary eccentricity growth it is absolutely necessary to consider viscous spreading of the circumbinary disk.

\begin{figure}
\centering
\includegraphics[width=0.5\textwidth]{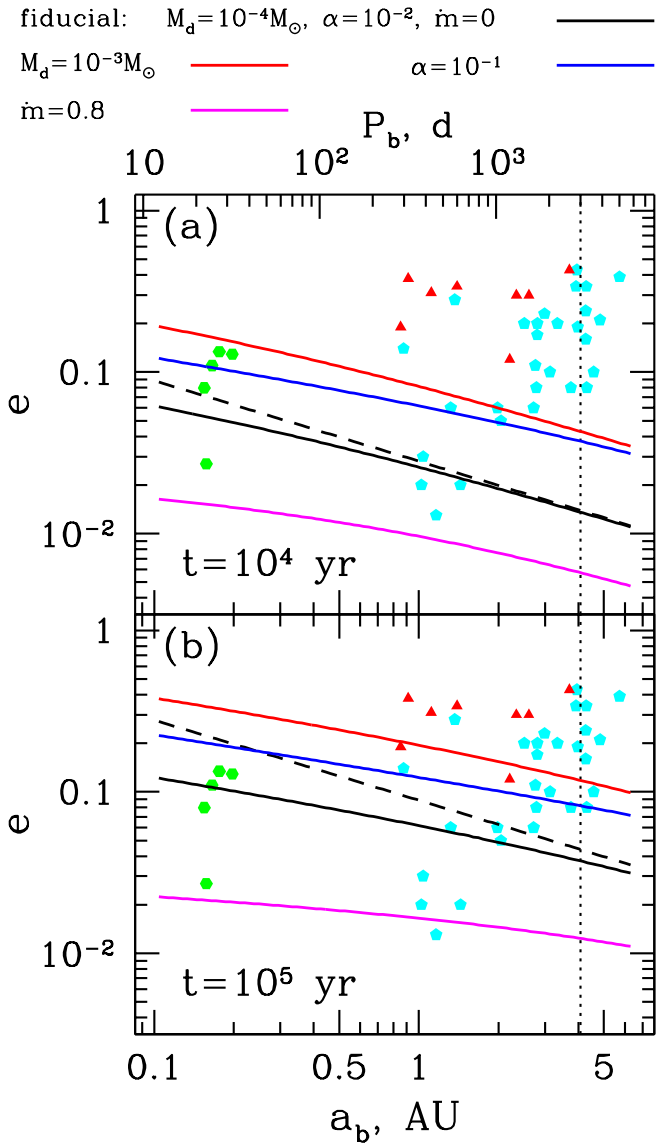}
\caption{Eccentricity of the binary accumulated due to the gravitational coupling to the disk by (a) $t=10^4$yr and (b) $10^5$yr, as a function of the binary semi-major axis $a_b$ (corresponding binary period $P_b$ is shown on top axis). Binary with $M_b=M_\odot$ and $q=0.3$ is assumed throughout. Black curves correspond to the fiducial disk with the parameters indicated on top. Other curves result from variation of one of the parameters as indicated in the legend. Black dashed curves are for the fiducial set of parameters with the disk torque kept at its initial level (i.e. disregarding circumbinary disk evolution). Data points are shown for (green diamonds) anomalously eccentric millisecond pulsars with the white dwarf companions and $P_b\sim 30$ d  \citep{Antoniadis2016}, (red triangles) post-AGB binaries \citep{vanWinckel}, and (cyan pentagons) strong Ba stars \citep{Jorissen}. Vertical dotted line corresponds to $P_b=3000$ d, out to which these binaries are expected to be circularized according to \citet{Pols}.
\label{fig:e_f_a_b}}
\end{figure}


\section{Discussion}  
\label{sect:disc}


We now compare our results on the eccentricity excitation with the observed orbital parameters of the post-MS binaries. In Figure \ref{fig:e_f_a_b} we show the eccentricity reached by the systems with different parameters according to our calculations, at two different evolution times, as a function of the binary semi-major axis $a_b$. Also plotted are the data for several classes of the eccentric post-MS systems --- binary MSPs with the WD companions and $P_b\sim 30$ d, post-AGB stars, and Ba stars.

Our fiducial system has relatively light disk ($M_d=10^{-4}M_\odot$) and $\alpha=10^{-2}$. Rather optimistically we also set $\dot m=0$, i.e. assume disk mass to be preserved during its viscous evolution (black solid curve). One can see that in this case we can account for the eccentricities of the high-$e$ binary MSPs with $P_b\sim 30$ d if the disk persists for $10^5$ yr. However, this set of parameters can explain eccentricities of only a small number of the Ba stars, and none of the post-AGB stars, even if the disk stays around for $10^5$ yr, see Figure \ref{fig:e_f_a_b}b. Moreover, as soon as reaccretion is allowed (see the magenta curve for $\dot m=0.8$) even the eccentricities of the MSP+WD binaries become difficult to account for. 

Figure \ref{fig:e_f_a_b} also demonstrates that for a given set of the binary+disk parameters the efficiency of eccentricity excitation {\it decreases} with $a_b$. This is naturally explained by the slower initial viscous evolution of the disk for wider binaries, see equation (\ref{eq:t0}). However, eccentricities of the  observed systems clearly do not conform to this trend, even though it might be not easy to establish a clear correlation in the $a_b-e$ space for different classes of the post-MS binaries (one has to also keep in mind the steep  $P_b$-dependence of the tidal circularization during the AGB phase). In his study of the MSP+WD binaries \citet{Antoniadis} found the efficiency of eccentricity excitation to rise with $a_b$; however, this was achieved by having a steep dependence of the disk lifetime on $a_b$, $t\propto a_b^2$, which we do not consider in our work.

Even boosting the disk mass by an order of magnitude ($M_d=10^{-3}M_\odot$, red curves) still does not help to explain eccentricities of most of the post-AGB and Ba star-hosting systems with $P_b\lesssim 3000$ d, which should have been circularized during the AGB phase \citep{Pols}. To account for the high $e$ of most of the objects one would need several conditions to be fulfilled {\it simultaneously}:
\begin{itemize}
\item a long-lived ($\gtrsim 10^5$ yr) circumbinary disk, 
\item which starts out rather massive ($M_d\gtrsim 10^{-2}$), \item and  does not reaccrete onto the binary during its viscous evolution ($\dot m=0$). 
\end{itemize}
It is not obvious that these conditions can be naturally satisfied simultaneously, as we discuss now.

The lifetimes of the disks around evolved binaries are highly uncertain but estimated to be $\lesssim 10^5$ yr \citep{Dermine}. However, this uncertainty is unlikely to strongly impact our conclusions since binary eccentricity grows very slowly with time (clearly visible in Figure \ref{fig:e_f_a_b}) --- roughly as $t^{1/4}$ for the case of no reaccretion, see equation (\ref{eq:et}). Eccentricity growth is even slower when reaccretion is allowed: combining equations (\ref{eq:L_devolution}) and (\ref{eq:master}) one can see that in the low-$e$ limit $e(t)\propto t^{(1-k_\varphi)/2}$, and $k_\varphi\to 1$ as the reaccretion efficiency increases (i.e. as $\dot m\to 1$ and $\lambda\to 1/2$); note the very small change of $e$ at short periods in the $\dot m=0.8$ case between the two panels of Figure \ref{fig:e_f_a_b}.

Circumbinary disk masses are also rather uncertain. For MSPs $M_d$ is inferred to be about $10^{-4}-10^{-3}M_\odot$ from the stellar evolution calculations, as the mass ejected by the WD progenitor as it undergoes H-shell flashes \citep{Antoniadis}. It is not guaranteed that all of this mass will remain bound to the binary in a stable Keplerian disk rather than being ejected to infinity; thus, the actual value of $M_d(0)$ could be lower. In the case of post-AGB binaries, we have some idea of the masses of their circumbinary disks based on the molecular (CO) line observations. A number of evolved systems show line emission consistent with $\sim 10^{-3}-10^{-2}M_\odot$ locked in a gas reservoir around the binary \citep{BujarrabalDisks}. However, only in two systems --- Red Rectangle \citep{BujarrabalRedRect} and AC Her \citep{BujarrabalACHer} --- we see clear evidence of the Keplerian rotation in the circumbinary gas reservoir based on interferometric measurements, implying stable bound disks. In both cases disk masses are $\lesssim 10^{-2}M_\odot$.


\subsection{Effect of reaccretion}  
\label{sect:reaccr}


Probably the most critical issue for the whole idea of the disk-driven eccentricity excitation in the post-AGB binary is that of reaccretion of gas from the inner disk back onto the binary. Reaccretion reduces the torque acting on the binary \citep{Rafikov2016}, slowing down eccentricity evolution. This can be seen in Figure \ref{fig:time_ev}c as the lower $e$ reached by the system with reaccretion ($\dot m=0.45$, black) compared to the case of the completely suppressed accretion ($\dot m=0$, magenta). This is not surprising since in the former case the torque acting on the binary is lower in amplitude and decays faster than in the latter, see the corresponding curves in Figure \ref{fig:time_ev}b.

Figure \ref{fig:e_f_a_b} demonstrates that reaccretion at the rate of $80\%$ of the $\dot M$ that a disk with no central torque (a standard accretion disk assumption) would have ($\dot m=0.8$, magenta curve), results in the reduction of $e$ by a factor of $3-5$ compared to the case of no reaccretion. This difference grows only larger with time, because of slower growth of $e$ in disks with reaccretion, as discussed above. 

\begin{figure}
\centering
\includegraphics[width=0.5\textwidth]{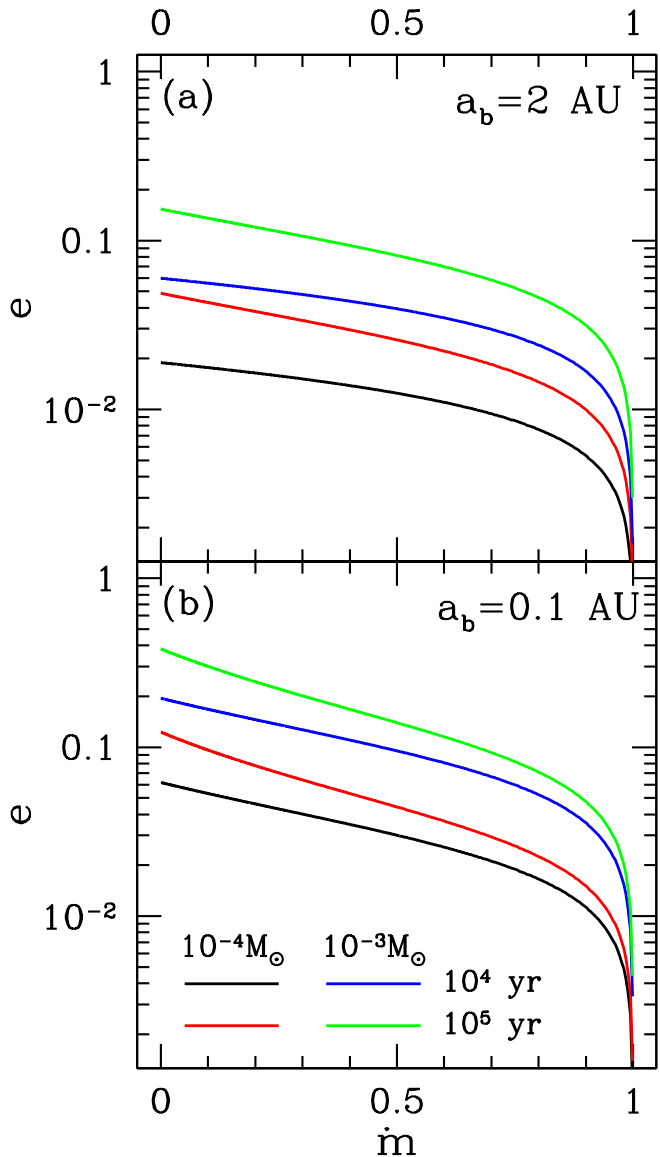}
\caption{Dependence of the final binary eccentricity on the degree of reaccretion $\dot m$. Values of $e$ are shown for two systems with (a) $a_b=2$AU and (b) $a_b=0.1$AU, and for different values of the initial disk mass ($10^{-4}M_\odot$ and $10^{-3}M_\odot$) and its lifetime ($10^4$yr and $10^5$yr) as indicated on the Figure. 
\label{fig:e_f_mdot}}
\end{figure}

Figure \ref{fig:e_f_mdot} illustrates the role of reaccretion in more detail, by showing $e$ reached by $t=10^4$ yr and $10^5$ yr in different systems as a function of $\dot m$. One can see that $e$ is a rapidly decreasing function of $\dot m$. For $\dot m=0.8$ one finds eccentricity, which is $\sim 3$ lower than that in systems with no reaccretion. Thus, understanding the behavior of gas near the inner edge of the circumbinary disk and accurately measuring $\dot m$ is very important for obtaining a reasonable estimate of the final binary eccentricity.

Numerical simulations of the circumbinary disks performed in studies of the binary supermassive black holes in centers of galaxies \citep{MacFadyen2008,DOrazio2013} and young stellar binaries \citep{Munoz} invariably show that fluid at the inner edge of the disk has no trouble entering the central cavity and being accreted by the binary. Moreover, simulations using different methods \citep{Farris2014,ShiKrolik2015,Munoz} typically show that reaccretion is only weakly (if at all) suppressed by the binary torque for disks with aspect ratio $h/r\sim 0.1$. These simulation results imply that $\dot m\to 1$, which, according to Figure \ref{fig:e_f_mdot}, makes eccentricity excitation by tidal coupling to the disk pretty much hopeless: even for the highest disk masses one would find $e\lesssim 0.1$ for post-AGB stars with $P_b\gtrsim 1$ yr, in clear disagreement with the observational data shown in Figure \ref{fig:e_f_a_b}. Reaccretion may be less efficient if the disk is colder \citep{Ragusa}, but circumbinary disks around post-AGB binaries with $P_b\gtrsim 1$ yr should have $h/r\gtrsim 0.1$ anyway.

Possibility of reaccretion in post-AGB binaries is also supported by the empirical evidence. Many of these systems show strong depletions of refractory elements, e.g. Fe, in their post-AGB component \citep{vanWinckel1995}. This was interpreted as evidence for reaccretion by \citet{Waters} and \citet{Maas}, who suggested that gas ejected by the binary and locked in the circumbinary disk can cool and efficiently form dust, thus removing refractory elements from the gas phase. The metal-poor gas then gets reaccreted by the post-AGB star resulting in the observed depletion patterns. While this interpretation is probably non-unique, it does seem natural in light of the aforementioned numerical results. Interestingly, the byproduct of this process --- a disk enriched in refractory dust --- can provide fertile ground for the second generation planet formation, previously proposed \citep{Volschow} to explain the origin of the putative post-common envelope planetary systems such as NN Ser \citep{Marsh,NNSerDust}.  

One may wonder if reaccretion can be observed directly in post-AGB binaries, which (unlike the Ba stars and MSPs with the WD companions) still harbor circumbinary disks. In Figure \ref{fig:time_ev}a we show the time evolution of the central mass accretion rate onto the binary in different systems with reaccretion (i.e. $\dot m\neq 0$). One can see that relatively high accretion rates $\gtrsim 10^{-7}M_\odot$ yr$^{-1}$ for the  short-period systems ($a_b=0.1$AU, progenitors of the MSP+WD systems) with high-mass disks ($M_d\sim 10^{-3}M_\odot$) persist for rather short viscous time $\sim 10^3$ yr. Lighter and more evolved ($t>10^4$ yr) disks have $\dot M\lesssim 10^{-9}M_\odot$ yr$^{-1}$. Given the possibility of still ongoing intrabinary accretion due to the RLOF or wind, such as e.g. inferred in the Red Rectangle \citep{Witt}, directly measuring signatures of the circumbinary reaccretion is likely to be too challenging.


\subsection{Validity of assumptions}  
\label{sect:validity}


Here we justify some of the assumptions used in this work. Our study assumes effective viscosity $\alpha=10^{-2}$, while \citet{Antoniadis} and \citet{Dermine} adopted $\alpha=0.1$ in their calculations of the disk-driven binary eccentricity growth. As can be inferred from Figures \ref{fig:time_ev}c and \ref{fig:e_f_a_b}, this would result in the increase of the binary eccentricity by $\sim 2$, in rough agreement with the estimate (\ref{eq:et}). 

The higher value of $\alpha$ may be reasonable at the inner edge of a circumbinary disk around a compact binary ($a_b\sim 0.1$ AU), where the gas is well ionized. Then the conditions are similar to the disks of cataclysmic variables, which are inferred to have $\alpha\sim 0.1-0.4$ \citep{King}. However, in our calculations viscosity enters as the driver of the {\it global} disk evolution, and what matters is its value at the {\it outer} disk edge, where most of the mass is concentrated. As soon as the outer disk radius exceeds $\sim 1$ AU, dust formation becomes possible, making conditions in the spreading circumbinary disks more akin to the protoplanetary disks. The latter are known to be less viscous, $\alpha\lesssim 10^{-2}$ \citep{Hartmann}, justifying our adopted value of $\alpha$.

Throughout this work we have assumed that disk is heated by the central source with luminosity $L_\star=10^3L_\odot$. Such $L_\star$ is typical for the post-AGB stars with initial mass around $1M_\odot$, but more massive progenitors are more luminous on the post-AGB branch, with $L_\star\sim 2\times 10^4L_\odot$ for $7M_\odot$ progenitor \citep{vanWinckelARAA}. This would result in hotter disk, which would spread faster and absorb more angular momentum from the binary, potentially leading to higher $e$. However, such brighter post-AGB stars remain in their luminous state for only a short interval of time, $\lesssim 10^3$ yr, compared to $\sim 10^5$ yr for $1M_\odot$ progenitor. Moreover, as equation (\ref{eq:et}) demonstrates, binary eccentricity has extremely weak sensitivity to central luminosity, $e\propto L_\star^{1/16}$. Thus, the effect of higher luminosity on the disk evolution is very short-lived and should not affect our results. 

Finally, in our calculations we neglected the evolution of the semi-major axis of the binary, which must be driven by the same tidal torques that excite its eccentricity. First, as shown by \citet{Dermine}, evolution of $a_b$ is very slow, and our results depend on $a_b$ very weakly anyway, see equation (\ref{eq:et}). Second, using \citet{Lubow1996} one can show that in the low-$e$ limit the binary interacting with the circumbinary disk mainly via its 2:1 resonance should lose roughly equal amounts of angular momentum to shrinking its semi-major axis and increasing its eccentricity.  Thus, our simplifying assumption that {\it all} torque acting on the binary goes into increasing its eccentricity leads to an {\it overestimate} of $e$ (by $\sim\sqrt{2}$). As a consequence, the prospects for driving binary eccentricity via disk torques are even more pessimistic than our calculations imply.


\subsection{Comparison with the existing work}  
\label{sect:compar}


Our work is different from the existing calculations of the binary eccentricity excitation by \citet{Dermine} and \citet{Antoniadis} in three important ways. First, as previously discussed, we recognize the need to lower the effective viscosity $\alpha$ when calculating disk evolution, which directly translates into the central torque calculation.

Second, we fully account for the viscous expansion of the disk, which, as we argue, is a necessary ingredient for increasing the binary eccentricity. Spreading disk provides a growing sink for the angular momentum that the binary can lose, and determines the rate at which this process happens. The fact that the instantaneous global angular momentum flux at the disk center determines the strength of the resonant torque between the disk and the binary was well recognized by \citet{Lubow1996} and \citet{Lubow2000}. However, in subsequent applications the importance of the evolution of the global viscous angular momentum flux was overlooked, resulting in a significant overestimate of the binary eccentricity. 

This is illustrated in Figures \ref{fig:time_ev} and \ref{fig:e_f_a_b}, where the dashed curves correspond to systems with our fiducial set of parameters but evolved as if the disk was not changing, effectively keeping its surface density at $\Sigma\sim M_d(0)/a_b^2$ (as implicitly assumed by previous authors) at all times; in practice we just fix the disk torque at its value at $t=0$ in these calculations. One can clearly see the faster growth of eccentricity ($e\propto t^{1/2}$) in the constant torque case, resulting in the overestimate of $e$ at $t=10^5$ yr compared to what our calculations yield.

Third modification is the recognition of the role of reaccretion for the disk evolution and for the binary eccentricity evolution. By not evolving the disk \citet{Dermine} and \citet{Antoniadis} effectively disregarded reaccretion, and their disk mass was unchanged while the binary eccentricity kept growing. But as we show here, even modest degree of reaccretion results in significant reduction of the eccentricity excitation, see Figure \ref{fig:e_f_mdot} and \S \ref{fig:e_f_mdot}. Thus, it is very important both to understand the amplitude of reaccretion in post-MS binaries and to properly account for it.


\section{Summary}  
\label{sect:sum}


We explored the excitation of eccentricities in the post-MS binaries by their tidal coupling to a circumbinary disk, produced by the initial episode of mass ejection from the binary. Compared to the existing studies, we include several novel ingredients, while keeping track of the global conservation of the angular momentum. First, we fully account for the viscous spreading of the circumbinary disk and the associated evolution of the disk-binary torque. Second, we consider the possibility of reaccretion of material from the disk by the binary, in agreement with the recent numerical simulations and empirical evidence in post-AGB binaries. Both processes act to lower the surface density at the inner edge of the circumbinary disk, thus reducing the efficiency of the disk-binary coupling. Moreover, we argue that truly global models of this coupling should employ lower value of effective viscosity ($\alpha\sim 10^{-2}$) than previously assumed. 

All these factors conspire to result in a considerably weakened disk-driven excitation of the eccentricities in the post-MS binaries, compared to previous studies. While we can account for the observed eccentricities ($e\approx 0.03-0.13$) of the MSP+WD binaries with $P_b\sim 30$ d with our standard set of parameters and no reaccretion, even modest degree of reaccretion is enough to destroy the agreement. Situation is even worse for post-AGB binaries and Ba stars as the high eccentricities of many of them (up to 0.4-0.5) would require massive ($\gtrsim 10^{-2}M_\odot$), long-lived ($\gtrsim 10^5$ yr) circumbinary disks that do not reaccrete. We find this combination (especially its last aspect) rather non-trivial, concluding that eccentricities of the post-MS binaries are unlikely to be excited by the tidal coupling to the circumbinary disk.  

Our results motivate exploration of other mechanisms that could be responsible for the high eccentricities of the evolved binaries with $P_b\lesssim$ several yr.

\acknowledgements

I am grateful to Ben Bar-Or for bringing the puzzle of the eccentricities of the MSPs with the WD companions to my attention. R.R.R. is an IBM Einstein Fellow at the IAS. Financial support for this study has been provided by the NSF via grants AST-1409524,  AST-1515763, NASA via grants 14-ATP14-0059, 15-XRP15-2-0139, and The Ambrose Monell Foundation.


\bibliographystyle{apj}
\bibliography{references}

\end{document}